\documentclass[letterpaper, 10 pt, conference]{ieeeconf}  

\IEEEoverridecommandlockouts                              

\usepackage{graphicx}
\usepackage{epstopdf}
\usepackage{multicol}
\usepackage{stfloats}
\usepackage{amsfonts}

\usepackage{mathrsfs}
\usepackage{multicol}
\usepackage{stfloats}
\usepackage{cite}
\usepackage{amsfonts}
\usepackage{mathrsfs}
\usepackage{amsfonts}
\usepackage{empheq}
\usepackage[linesnumbered,ruled,lined,boxed]{algorithm2e}
\usepackage[procnames]{listings}
\usepackage{algpseudocode}
\usepackage{color}
\usepackage[ruled]{algorithm2e}
\usepackage{fancybox}
\usepackage{framed}
\usepackage{amsfonts, color}
\usepackage{color}
\usepackage{setspace}
\usepackage[dvips]{epsfig}
\usepackage{framed}
\usepackage{psfrag, amssymb, amsmath, cite}
\usepackage[colorlinks,linkcolor=red,anchorcolor=blue,citecolor=blue]{hyperref}
\DontPrintSemicolon

\usepackage{verbatim}

\newtheorem{remark}{Remark}
\newtheorem{definition}{\textbf{Definition}}

\newtheorem{theorem}{\textbf{Theorem}}

\newtheorem{lemma}{\textbf{Lemma}}

\title{\LARGE \bf
Feasibility and coordination of multiple mobile vehicles with mixed equality and inequality constraints
}

\author{Zhiyong Sun, Marcus Greiff, Anders Robertsson and Rolf Johansson
\thanks{*The research leading to these results has received funding from
the Swedish Science Foundation (SSF) project ``Semantic mapping
and visual navigation for smart robots'' (RIT15-0038). }
\thanks{The authors are members of the LCCC Linnaeus Center and
the ELLIIT Excellence Center at Lund University, Sweden. Emails: 
        {\tt\small \{zhiyong.sun, marcus.greiff, rolf.johansson, anders.robertsson\} @control.lth.se}}%
}

\begin{document}

\maketitle
\thispagestyle{empty}
\pagestyle{empty}

\begin{abstract}
We consider the problem of feasible coordination control for multiple homogeneous or heterogeneous mobile vehicles subject to various constraints (nonholonomic motion constraints, holonomic coordination constraints, equality/inequality constraints etc). We develop a general framework involving differential-algebraic equations and viability theory to describe and determine coordination feasibility for a coordinated motion control under heterogeneous vehicle dynamics and different types of coordination constraints. 
If a solution exists for the derived differential-algebraic equations and/or inequalities, a heuristic algorithm is proposed for generating feasible trajectories for each individual vehicle. In case studies on coordinating two vehicles, we derive analytical solutions to motion generation for two-vehicle groups consisting of car-like vehicles, unicycle vehicles, or vehicles with constant speeds, which serve as benchmark coordination tasks for more complex vehicle groups. We show several simulation experiments on multi-vehicle coordination under various constraints to validate the theory and the effectiveness of the proposed schemes. 

\end{abstract}

\section{Introduction}
In the active research field of mobile robot motion planning and control, multi-vehicle coordination and cooperative control has been and will remain an attractive research topic, motivated by an increasing number of practical applications requiring multiple robots or vehicles to cooperatively perform coordinated tasks \cite{knorn2016overview,dames2017detecting,schwager2017multi}. These include multi-robot formation control, area coverage and surveillance, coordinated target tracking, to name a few \cite{oh2015survey,liu2018distributed}. A fundamental problem in multi-vehicle coordination is to plan feasible motion schemes and trajectories for each individual vehicle which should satisfy both kinematic or dynamic requirement for all vehicles, and inter-vehicle geometric constraints that describe the nature of a given coordination task. Typically, an individual vehicle is subject to various kinematic motion constraints which limit possible motion directions. A coordinated motion to achieve a predefined coordination task then further imposes inter-vehicle motion constraints, which makes the coordination control a challenging problem.  

The seminal paper by Tabuada \textit{et al.} \cite{tabuada2005motion} firstly studied the motion feasibility problem in the context of  multi-agent formation control. Via the tools of differential geometry, feasibility conditions were derived for a group of mobile agents to maintain formation specifications (described by strict equality constraints) in each agent's motions. Recently, the motion feasibility problem in multi-vehicle formation and cooperative control has resumed its interests in the control and robotics community. The paper \cite{maithripala2011geometric} discusses coordination control with dynamically feasible vehicle motions, and solves a rigid formation shape maintenance task and formation reconfiguration problem. 
Our recent work  \cite{Sun2016feasibility} investigates the formation and coordination feasibility with heterogeneous systems modelled by control affine nonlinear systems with drift terms (which include fully-actuated systems, under-actuated systems, and non-holonomic vehicles). More recently, the work by Colombo and Dimarogonas  \cite{colombo2018motion} extends the motion feasibility condition in \cite{tabuada2005motion} to multi-agent formation control systems on Lie groups. Cooperative transport control using multiple autonomous vehicles can also be formulated as a motion feasibility problem, while in \cite{hajieghrary2017cooperative} the authors discussed cooperative transport of a buoyant load using two autonomous surface vehicles (ASVs) via a differential geometric approach. The ASV's dynamics are described by the standard unicycle-type equations with non-holonomic constraints, while the two vehicles assume a cooperative task to maintain a fixed distance between them. 

Coordination tasks with mobile vehicles often involve  various types of inter-vehicle constraints, typically described by equality or inequality functions of inter-vehicle geometric variables. For example, a practical coordinated motion may be described by some inequality constraints that require a bounded inter-vehicle distance between mobile vehicles; i.e., a lower bound to guarantee collision avoidance, and an upper bound to avoid communication loss due to excessively long ranges. Furthermore, in multi-robotic visibility maintenance control, which requires vehicles' headings to lie in a bounded cone of field of view, coordination constraints are modelled by some inequality functions. All these practical coordination control scenarios call for a general framework for multi-vehicle coordination planning and control under various constraints. We remark that the above referenced papers \cite{tabuada2005motion,maithripala2011geometric,Sun2016feasibility,colombo2018motion,hajieghrary2017cooperative} only discussed formation or coordination control for multiple vehicles with \textit{strict equality} functions. This paper will focus on a more general problem in multi-vehicle coordination control that also includes inequality constraints, or a mix of equality and inequality constraints.  

The problem of maintaining holonomic equality constraints in multi-vehicle coordination is also relevant to the framework of virtual holonomic control (VHC). VHC involves a relation (usually described by an equality constraint) among the configuration variables of a mechanical or robotic system which does not physically exist \cite{freidovich2008periodic,    shiriaev2007virtual}. Such constraints are controlled invariant via feedback controllers \cite{maggiore2013virtual}. In this paper, we present a \textit{multi-vehicle} framework that includes equality constraints as a special case, and develop admissible control inputs that preserve both equality and inequality constraints. Our tools to solve feasible coordination problem of multiple vehicles with various constraints are an interplay of differential geometry for nonlinear control \cite{isidori1995nonlinear}, viability theory \cite{aubin2009viability} and differential-algebraic equations and inequalities. One of the key tools to address feasible coordination and motion generation with \textit{inequality} motion/coordination constraints is the viability theory \cite{aubin2009viability}, which has relevance in set-invariance control \cite{blanchini1999set} in the control theory (or termed controlled-invariance set).   It has been used in solving coordination control problem for under-actuated vehicles in \cite{panagou2014cooperative}, autonomous vehicle racing control in \cite{liniger2017real} and visibility maintenance for multiple robotic systems in \cite{panagou2013viability}.

In this paper, a synthesis of coordination control that respects vehicles' kinematic constraints (often modelled by nonholonomic motion constraints) and inter-vehicle constraints (which include holonomic formation constraints, inequality functions or a mix of various constraints) will be provided. We will also devise a heuristic algorithm to solve the proposed feasibility  equations and inequalities  that generate feasible trajectories for all vehicles to achieve a coordination task. We will consider two typical modellings for multiple vehicle coordination control, one based on undirected graph and the other based on leader-follower framework. In both cases we present feasibility conditions for vehicle coordination;  feasible motions and vehicle trajectories, if they exist,  can be generated by the devised heuristic algorithm. To illustrate the proposed coordination framework and theory, we also present several application examples and cases studies on coordinating two or more vehicles of homogeneous or heterogeneous kinematics, with equality or inequality coordination constraints between inter-vehicle distances or headings. We derive analytical solutions to motion generation for two-vehicle groups consisting of car-like vehicles, unicycle vehicles, or vehicles with constant speeds, which serve as benchmark coordination tasks for more complex vehicle groups.

This paper is organized as follows. Section~\ref{Sec:preliminary} provides preliminary knowledge of differential geometry, distribution/codistribution and introduces vehicle models. In Section~\ref{sec:constraints_formulation} we formulate motion constraints arising from vehicles' kinematics and coordination tasks in a unified way. Section~\ref{sec:coordination_feasibility} presents two key theorems to determine coordination feasibility and presents a heuristic algorithm for trajectory generation for the overall vehicle group. Case study and application examples on coordinating two or more homogeneous and heterogeneous vehicles are shown in Section~\ref{sec:case_study} 
(more results and demonstrations are shown in the accompanying video). Concluding remarks in Section~\ref{sec:conclusions} close this paper. 

\section{Preliminary and problem formulation} \label{Sec:preliminary}
In this section we introduce some standard notions and tools of differential geometry and nonlinear control systems from \cite{nijmeijer1990nonlinear, murray1994mathematical} which will  frequently be used in the main part of this paper. 

\subsection{Distribution, codistribution and vehicle models}
A distribution  $\Delta(x)$ on $\mathbb{R}^n$ is an assignment of a linear subspace
of $\mathbb{R}^n$ at each point $x$. Given a   set of $k$ vector fields $X_1(x), X_2(x), \cdots, X_k(x)$, we define the distribution as $$\Delta(x) = \text{span}\{X_1(x), X_2(x), \cdots, X_k(x)\}.$$ A vector field $X$ belongs to a distribution $\Delta$ if $X(x) \in \Delta(x)$, $\forall x \in \mathbb{R}^n$, and we assume all distributions have constant rank.

A codistribution assigns a subspace to the dual space, denoted by $(\mathbb{R}^{n})^\star$. Given a distribution $\Delta$,  for each $x$ consider the annihilator of $\Delta$, which is the set of all covectors that annihilates all vectors in   $\Delta(x)$
(see \cite[Chapter 1]{isidori1995nonlinear})
 $$\Delta^\perp = \{\omega \in  (\mathbb{R}^{n})^\star  | \left \langle \omega,   X \right \rangle = 0, \,\, \forall X \in \Delta \}.$$

In this paper, we model each individual vehicle's dynamics   by the following general form (i.e., control-affine system)
\begin{align} \label{eq:system_drift}
\dot p_i = f_{i,0} + \sum_{j=1}^{l_i} f_{i,j} u_{i,j},
\end{align}
where $p_i \in \mathcal{C}_i \in \mathbb{R}^{n_i}$ is the state of vehicle $i$ ($\mathcal{C}_i$ denotes the configuration for vehicle $i$, for which  we embed $\mathcal{C}_i$ in  $\mathbb{R}^{n_i}$ where $n_i$ denotes the dimension of state space for vehicle $i$), $f_{i,0}$ is a smooth drift term, and $u_{i,j}$ is the \emph{scalar} control input associated with the smooth vector field $f_{i,j}$, and $l_i$ is the number of vector field functions.
Such a nonlinear control-affine system~\eqref{eq:system_drift} with a drift term is very general in that it describes many different types of real-life vehicle dynamics and control systems, including control systems subject to under-actuation or nonholonomic motion constraints.

\subsection{Viability theory and set-invariance control}
In this paper, we will treat coordination tasks with inequality constraints, and a key tool to address inequality constraint is the viability theory and set-invariance control \cite{blanchini1999set,aubin2009viability}. We now introduce some background knowledge, concepts and theorems on viability theory. 

\begin{definition}
(\textbf{Viability and viable set}) Consider a control system described by a differential equation $\dot x(t) = f(x(t), u(t))$. A subset $\mathcal{F}$ enjoys the viability property for the system $\dot x(t)$  if for every initial state $x(0) \in \mathcal{F}$, there exists at least one solution to the system
starting at $x(0)$ which is viable in the time interval $[0, \bar t\,]$ in the sense that $$\forall t \in [0, \bar t \,], x(t) \in \mathcal{F}.$$
\end{definition}

We assume the solution of the differential system $\dot x(t) = f(x(t), u(t))$, modeling vehicle control systems under constraints, is well defined. 
When a differential equation involves discontinuous right-hand side (e.g., switching controls), we understand its solutions in the sense of Filippov \cite{cortes2008discontinuous}. 

Now define a distance function for a point $y$ to a set $\mathcal{F}$ as $d_\mathcal{F}(y)=:  \inf\limits_{z \in \mathcal{F}} \|y-z\|$, and consider the definition of contingent cone as follows. 
\begin{definition}
(\textbf{Contingent cone}) Let $\mathcal{F}$ be a nonempty subset of $\mathcal{X}$ and $x$ belongs to $\mathcal{F}$. The
contingent cone to $\mathcal{F}$ at $x$ is the set
\begin{align}
T_\mathcal{F}(x) = \left\{v \in \mathcal{X} |\,\,\,\,  \liminf\limits_{h \rightarrow 0^+} \frac{d_\mathcal{F}(x +hv)}{h} =0  \right\} 
\end{align}
\end{definition}
It has been shown in \cite{blanchini1999set} that though the distance function $d_\mathcal{F}(y)$ depends on
the considered norm, the set $T_\mathcal{F}(x)$ does not. Furthermore, the set $T_\mathcal{F}(x)$ is non-trivial only on the boundary of $\mathcal{F}$. 

A key result in the set-invariance analysis, the celebrated Nagumo theorem, is stated as follows (see \cite{blanchini1999set} or \cite{aubin2009viability}).
\begin{theorem}  \label{Theorem_Nagumo}
(\textbf{Nagumo theorem}) Consider the system
$\dot x (t) = f (x(t))$, and assume that, for each initial condition in
a set $\mathcal{X} \subset \mathbb{R}^n$, it admits a globally unique solution. Let $\mathcal{F} \subset \mathcal{X}$ be
a closed and convex set. Then the set $\mathcal{F}$ is positively
invariant for the system if and only if
\begin{align} \label{eq:Nagumo1}
f (x(t)) \in T_\mathcal{F}(x), \,\,\, \forall x \in \mathcal{F}. 
\end{align}
where $T_\mathcal{F}(x)$ denotes the \textit{contingent cone} of $\mathcal{F}$ at $x$. 
\end{theorem}

Generalizations of the Nagumo theorem and viability theory are also possible, by using the set-valued analysis and differential inclusion \cite{aubin2009set}.

If $x$ is an interior point in the set $\mathcal{F}$, then $T_\mathcal{F}(x) = \mathbb{R}^n$. Therefore, the condition in Theorem \ref{Theorem_Nagumo} is only meaningful when $x \in \text{bnd}(\mathcal{F})$, where $\text{bnd}(\mathcal{F})$ denotes the boundary of $\mathcal{F}$. Therefore, the condition in \eqref{eq:Nagumo1} can be equivalently stated 
\begin{align}
f (x(t)) \in T_\mathcal{F}(x), \,\,\, \forall x \in \text{bnd}(\mathcal{F}). 
\end{align}
The above condition clearly has an intuitive and geometric interpretation: if at $x \in \text{bnd}(\mathcal{F})$, the derivative
$\dot x   = f(x(t))$ points inside or  is tangent to $\mathcal{F}$, then the
trajectory $x(t)$ remains in $\mathcal{F}$. 

Now we consider a viable set $\mathcal{F}$ parameterized by an inequality associated with a continuously differentiable function $g(x): \mathbb{R}^n \rightarrow \mathbb{R}$,
\begin{align} \label{eq:set_K}
\mathcal{F} = \{x | g(x) \leq 0\}.
\end{align}
In this way, the calculation of $T_\mathcal{F}(x)$ is simplified to be
\begin{align} \label{viable_condition_inequality}
T_\mathcal{F}(x) = \{ v \in x | \left \langle v, \nabla g(x)\right \rangle \leq 0 \},
\end{align}
for any $ g(x) = 0$  and $T_\mathcal{F}(x) = \mathbb{R}^n$ when $g(x) < 0$. For the   set  $\mathcal{F}$ defined in \eqref{eq:set_K}, a consequence of Nagumo theorem is the following lemma on a controlled-invariant set. 
\begin{lemma} \label{lemma:invariant_set_control}
(\textbf{Set-invariance in control}, \cite{blanchini1999set})  Consider a set $\mathcal{F}$ parameterized by an inequality of a continuously differentiable function $g(x)$: $\mathcal{F} = \{x| g(x) \leq 0\}$. Then the set $\mathcal{F}$ is positively invariant under the dynamic control system $\dot x(t) = f(x(t), u(t))$ if $\dot x(t) \in T_\mathcal{F}(x)$ of \eqref{viable_condition_inequality}, or equivalently 
\begin{align}
\left< \nabla g(x),  f(x(t), u(t))  \right> \leq 0, \,\,\, \forall x:  g(x(t)) = 0. 
\end{align}
\end{lemma}

\subsection{Problem formulation}

 Consider a group of $n$ vehicles, whose kinematic equations are described by the control-affine systems \eqref{eq:system_drift} with possibly different kinematics and/or drift terms. We assign the vehicle group with a coordination task, described by inter-vehicle geometric equality or inequality constraints that incorporate formation, flocking or other cooperative tasks. Two key problems to be addressed in this paper are the following:
 \begin{itemize}
 \item  Determine whether a group of homogeneous or heterogeneous vehicles can perform a coordination task with various constraints;
\item If the coordination task with various constraints is feasible, determine feasible motions that generate trajectories for an $n$-vehicle group to perform the task. 
 \end{itemize}

\section{Formulation of coordination constraints} \label{sec:constraints_formulation}
\subsection{Motion constraints arising from vehicle kinematics} \label{sec:constraint_kinematics}
In this subsection we follow the techniques in \cite{murray1994mathematical,Sun2016feasibility} to formulate vehicle's kinematic constraints using (affine) codistributions. 
A vehicle's kinematics modelled by a  nonlinear control-affine system \eqref{eq:system_drift} with drifts can be equivalently described by the following \textit{affine} distribution
\begin{align}
\Delta_i  = f_{i,0} + \text{span}\{f_{i,1}, f_{i,2}, \cdots, f_{i,l_i}\}.
\end{align}
For the system \eqref{eq:system_drift} with drifts, one can obtain a corresponding transformation with equivalent  constraints via the construction of covectors
\begin{align} \label{eq:drift_equi}
\omega_{i, j}(p_i) \dot p_i = q_{i, j},  \quad j =1, \cdots, n_i - l_i,
\end{align}
where  the term $q_{i, j}$ is due to the existence of the drift term $f_{i,0}$.  We collect all the \textit{row covectors} $\omega_{i, j}$ as $\Omega_{K_i} = [\omega_{i,1}^{\top}, \; \omega_{i,2}^{\top},\; \cdots,\; \omega_{i,n_i - l_i}^{\top}]^{\top}$, and similarly define $T_{K_i} = [q_{i,1}, q_{i,2}, \cdots, q_{i,n_i - l_i}]^{\top}$. By doing this, one can rewrite \eqref{eq:drift_equi} in a compact form as follows
 \begin{align} \label{eq:dynamics_constraint_transformation}
 \Omega_{K_i}  \dot p_i = T_{K_i},
\end{align}
where the subscript $K$ stands for \emph{kinematics}.
Furthermore, we collect all the kinematic  constraints for all the $n$ vehicles in a composite form
$$\Omega_K = [\Omega_{K_1}^{\top}, \Omega_{K_2}^{\top}, \cdots, \Omega_{K_n}^{\top}]^{\top}, T_K = [T_{K_1}^{\top}, T_{K_2}^{\top}, \cdots, T_{K_n}^{\top}]^{\top}.$$
 For ease of notation, we collect all of the vehicles' states together, denoting them by the composite state vector $P = [p_1^{\top}, p_2^{\top}, \cdots, p_n^{\top}]^{\top}$. Thus, the overall kinematic constraint for all the vehicles can be stated compactly as $\Omega_K(\dot P) = T_K$.
 
\begin{remark}
The kinematics of the drift-free vehicle model 
\begin{align} \label{eq:system_drift_free}
\dot p_i = \sum_{j=1}^{l_i} f_{i,j} u_{i,j}
\end{align}
can be described in an equivalent form
\begin{align} \label{eq:drift_free_equi}
\omega_{i, j}(p_i) \dot p_i = 0,  j =1, \cdots, n_i - l_i.
\end{align}
i.e.,  the term  $q_{i, j}$ becomes zero. The above transformation is based on the idea that  a distribution generated by vector fields of a nonlinear control system can be equivalently defined by its  annihilating codistribution \cite{nijmeijer1990nonlinear}. Note that each $\omega_{i, j}(p_i)$ in \eqref{eq:drift_free_equi} is a \emph{row} covector in the dual space $(\mathbb{R}^{n_i})^\star$.
\end{remark}

\subsection{Motion constraints arising from coordination tasks} \label{sec:coordination_constraints}
In this section we formulate motion constraints from coordination tasks using distributions/codistributions. We consider two types of constraints, \textit{equality} constraints and \textit{inequality} constraints, which both  involve inter-vehicle geometric relationships, in modelling a general form of coordination tasks. 

\subsubsection{Coordination with equality constraints}
In this section, we assume a networked multi-vehicle control system modelled by an undirected graph $\mathcal{G}$, in which we use   $\mathcal{V}$ to denote its vertex set and $\mathcal{E}$ to denote the edge set.  The vertices consist of   $n$ homogeneous or heterogeneous vehicles, each modelled by the general dynamical equation \eqref{eq:system_drift} with possibly different dynamics. The graph consists of  $m$ edges, each associated with one or multiple inter-vehicle constraints describing a coordination task.

A family of equality constraints $\Phi$ is indexed by the edge set, denoted as $\Phi_\mathcal{E} = \{\Phi_{ij}\}_{(i,j)}$ with  $ {(i,j)} \in \mathcal{E}$. For each edge $(i,j)$, $\Phi_{ij}$ is a continuously differentiable vector function of the states $p_i$ and $p_j$  defining the coordination constraints between the vehicle pair $i$ and $j$.
The constraint for edge $(i,j)$ is enforced if $\Phi_{ij} (p_i, p_j) = 0$.  Such equality constraints can be used to describe very general coordinate control problems, such as formation shape control, distance maintenance, tracking and coverage control. For example, in formation shape control, the constraint vector function $\Phi_{ij}$ can be functions of desired relative position, or desired bearings, or desired distances between vehicles $i$ and $j$ describing a target formation (see e.g., \cite{oh2015survey}).  To satisfy the equality constraint for edge $(i,j)$, it should hold that
\begin{align} \label{eq:constraints_ij}
\frac{\text{d}}{\text{d}t} \Phi_{ij} = \frac{\partial \Phi_{ij}}{\partial p_i} \dot p_i + \frac{\partial \Phi_{ij}}{\partial p_j} \dot p_j + \frac{\partial \Phi_{ij}}{\partial t} = 0.
\end{align}
 
 We collect the equality constraints for all the edges and define an overall  constraint denoted by $\Phi_\mathcal{E}  = [\cdots, \Phi_{ij}^{\top}, \cdots]^{\top} = \bf 0$. A coordination task is  maintained if $\Phi_\mathcal{E} (P) = 0$ is enforced for all the edges.  Coordination feasibility with equality constraints means that the constraints are strictly satisfied along the  trajectories of all vehicles in time. Thus, one can obtain
\begin{align} \label{eq:constraints_all}
\frac{\text{d}}{\text{d}t} \Phi =\frac{\partial \Phi}{\partial P} \dot P + \frac{\partial \Phi}{\partial t} = 0,
\end{align}

Now we group all the constraints for all the edges by writing down a compact form $ T_E = - [\cdots, (\frac{\partial \Phi_{ij}}{\partial t})^{\top}, \cdots]^{\top}$ and identify a codstribution matrix $\Omega_E$  associated with the Jacobian $\frac{\partial \Phi}{\partial P}$ using the nominal dual coordinate bases $\text{d}[P]$. 
We now can reexpress equation \eqref{eq:constraints_all} as
\begin{equation}
\Omega_E(\dot P) = T_E.
\end{equation}
where the subscript $E$ stands for \emph{equality} constraints. For time-invariant equality constraint, one has $T_E = 0$.  Thus, the vector field $\dot P$ defined by the above equation represents possible motions for all the vehicles that respect the coordination equality constraint.

\subsubsection{Coordination with inequality constraints}

Now we consider a feasible coordination problem involving \textit{inequality} constraints. A family of inequality coordination constraints $\mathcal{I}_\mathcal{E} = \{\mathcal{I}_{ij}\}_{(i,j)}$ is indexed by the edge set $\mathcal{E}$, and  each edge $(i, j)$ is associated with a vector function $\mathcal{I}_{ij}(p_i, p_j)$ which is assumed continuously differentiable. The constraints for the edge $(i, j)$ are enforced if $\mathcal{I}_{ij}(p_i(t), p_j(t)) \leq 0\;\;\forall t$. Now we consider the subset of \textit{active} constraints among all the edges
\begin{align}
\chi(P) = \{(i, j), \;i, j = 1,2, \cdots, n\; |\; \mathcal{I}_{ij}(p_i, p_j) = 0\}.
\end{align}
We remark that the set $\chi(P)$ is a dynamic set along time, which only collects the edge set with active constraints when the condition $\mathcal{I}_{ij}(p_i, p_j) \leq 0$ is about to be violated. For simplicity we consider time-invariant functions $\mathcal{I}_{ij}(p_i, p_j)$. An inequality constraint for edge $(i,j)$ is maintained if 
\begin{align}
    \frac{\text{d}}{\text{d}t} \mathcal{I}_{ij} = \frac{\partial \mathcal{I}_{ij}}{\partial p_i} \dot p_i + \frac{\partial \mathcal{I}_{ij}}{\partial p_j} \dot p_j \leq  0, \,\,\,\forall\,(i, j) \in \chi(P).
\end{align}
At any point in time, all the active constraints in the edge set $\chi(P)$ generate a codistribution
\begin{align}
\Omega_I = [\cdots, \Omega_{I, ij}^\top, \cdots]^\top, \,\,\,\forall (i, j) \in \chi(P),
\end{align}
where the   subscript $I$ stands for \emph{inequality} constraints, and   $\Omega_{I, ij}$ is obtained by the Jacobian of the vector function $\mathcal{I}_{ij}$ using the nominal coordinate bases $[\text{d}p_i, \text{d}p_j]$ associated with the active constraint $\mathcal{I}_{ij}(p_i, p_j) = 0$. 
Based on the Nagumo theorem and Lemma~\ref{lemma:invariant_set_control},  to guarantee the validity of the inequality constraints, the control input $u(t) = [u_1(t)^\top, \cdots, u_n(t)^\top]^\top$ for each vehicle should be designed such that $\Omega_I \dot P(P, u(t)) \leq 0$, $\forall (i, j) \in \chi(P)$. 

\section{Coordination feasibility and motion generation} \label{sec:coordination_feasibility}
\subsection{Coordination feasibility with inequality task constraints}

We now state the following theorem on a feasible coordination for an $n$-vehicle group with kinematic constraint and inequality constraints in a coordination task. 
\begin{theorem}\label{thm:feasible_equality}
The coordination task with inequality constraints has feasible motions if the following mixed (in)equalities have solutions
\begin{align}
\Omega_K \dot P &= T_K, \nonumber \\
\Omega_I \dot P  & \leq  0,\,\,\, \forall (i, j) \in \chi(P),
\end{align}
where $\chi(P)$ denotes the set of active constraints among all the edges. 
\end{theorem}

\begin{remark}
The expression of the codistribution   $\Omega_I$ of active inequality constraints is coordinate-free and is also independent of the enumeration of edge sets. However, one can always choose the nominal coordinate bases $[\text{d}P]$ to present the codistribution $\Omega_K$ and $\Omega_I$ in a matrix form. 
\end{remark}

\subsection{Coordination feasibility of multiple vehicles with both  equality and inequality task   constraints}
We now consider a coordination task with both equality and inequality constraints.  
Together with the active inequality constraints, one can state the following theorem that determines coordination feasibility with various constraints. 
\begin{theorem} \label{thm:equality_inequality}
The coordination task with both equality and inequality constraints has feasible motions if the following \textit{mixed} equations and inequalities have solutions
\begin{align}
\Omega_K \dot P &= T_K, \nonumber \\
\Omega_E \dot P &= T_E, \nonumber \\
\Omega_I \dot P  & \leq  0,\,\,\, \forall (i, j) \in \chi(P),
\end{align} 
where $\chi(P)$ denotes the set of active constraints among all the edges. 
\end{theorem}

The above theorem is a generalization of the main result of \cite{tabuada2005motion} which derived a feasibility condition for multi-agent formation with only equality constraints. 
Again, the expression of the codistribution $\Omega_E$ and $\Omega_I$ is coordinate-free and is also independent of the enumeration of edge sets. One can present them in a matrix form using the bases $[\text{d}P]$ of the dual space for the convenience of calculations. 

\subsection{Generating vehicle's motion and trajectory for a feasible coordination}
The feasibility conditions presented in Theorems \ref{thm:feasible_equality} and~\ref{thm:equality_inequality} involve the determination of the existence of solutions for an algebraic equation (or a mixed inequality with equations). Solving these   equations with inequalities also leads to feasible motions that generate trajectories for each individual vehicle that meets both its own kinematic/dynamic constraints and the inter-vehicle constraints for performing a coordination task. Generally speaking, when a solution exists that meets the differential-algebraic equations/inequalities, then such a solution is not unique. Any feasible trajectories can be generated by possible motions as described by the solutions of these equations/inequalities. 

We remark that available approaches in numerical differential geometry and nonlinear control (see e.g., \cite{kwatny2000nonlinear}) are helpful and can be employed in solving these algebraic equations/inequalities. Furthermore, certain commercial software 
(e.g., \textit{Matlab} or \textit{Mathematica}) has powerful toolboxes available that can perform symbolic computations if the number of symbolic variables is within a reasonable scale. They provide an alternative approach for solving the equations/inequalities in the theorems that generate admissible trajectories for a feasible coordination. 

Algorithm \ref{algorithm:feasibility_undirected} presents a heuristic approach to determine coordination feasibility and motion generation for the multi-vehicle coordination control under both equality and inequality constraints. When a feasible motion is determined with a set of virtual input $w_l$, the actual control input $u_i$ can be readily calculated via each vehicle's kinematic equations. 

\begin{algorithm}[t!]
 Initialization: $\Omega_{K_i}$, $T_{K_i}$, $\Omega_{E}$, $T_E$, $\chi(P)$, $\Omega_I$;\;
 Construct the overall kinematic codistribution matrix~$\Omega_K$ and the vector $T_{K}$.\;

 \BlankLine
 \While{Running}{
 \BlankLine
  \textit{Solve equality} $ \left[ \begin{array}{c}
\Omega_K\\
 \Omega_E   
  \end{array} \right]  \dot P=\left[ \begin{array}{c}
T_K \\
T_E   
  \end{array} \right]
$\\
 \uIf{Solution does not exist}{
   Return: No solution;\;
   Condition checking STOP.\;
 }
 \Else{
    Calculate a special solution to the above equality constraint equation, denoted by $\bar K$;\;
    Determine $\kappa$ vectors of $\text{Null}\left(\left[ \begin{array}{c}
\Omega_K\\
 \Omega_E   
  \end{array} \right]\right)$, denoted by $K_1, K_2, \cdots, K_\kappa$.\;
  }
 \BlankLine
\uIf{$\chi(P) = \emptyset$ (No active inequality constraint) }{
 Feasible motions $\dot P = \bar K + \sum_{l=1}^\kappa K_l  w_l$, where $w_l$ is a set of virtual inputs that activate the associated vector field  $K_l$;\;
 Return: a set of feasible motions $\dot P = \bar K + \sum_{l=1}^\kappa K_l  w_l$ (according to different choices of $ w_l$).\;
}
\Else{
\For{$l=1, 2, \cdots, \kappa$}{
  Calculate and obtain the codistribution matrix $\Omega_I$ for active equality constraints, with $\forall (i, j) \in \chi(P)$;\;
 \If{$\Omega_I (\bar K + K_l  w_l)    \leq  0$ for certain $ w_l$ }{ 
  Return: A feasible motion $\dot P = \bar K +   K_l  w_l$. \;
  }
  }
  }
 \If{$l > \kappa$}{Return: Feasible solution not found. Try again.} 
 }
\caption{Coordination  feasibility checking and motion generation.}
\label{algorithm:feasibility_undirected}
\end{algorithm}

\section{Feasible coordination in a leader-follower vehicle group}
In this section we extend the above results to a leader-follower vehicle framework. Leader-follower structure involves a directed tree graph that describes the interaction relation within each individual vehicle, and has been used as a typical and benchmark framework in multi-vehicle coordination control (see e.g., \cite{panagou2013viability}). 

In a leader-follower structure, each vehicle has only one leader (and one or multiple followers). For each (directed) edge $(i, j)$ we associate a vector  function $\Phi_{ij}(p_i, p_j)$ to describe equality constraints. Different to the undirected graph modelling in the previous section,  here only the follower vehicle $j$ is responsible to maintain the equality constraint   $\Phi_{ij} = 0$ associated with edge $(i,j)$, and the leader vehicle $i$  is not affected by the equality constraint $\Phi_{ij}$. 

The equality constraint $\Phi_{ij}$ for edge $(i, j)$ is enforced along the trajectories of vehicles $i$ and $j$ if and only if $\Phi_{ij}(0) = 0$ and $\dot \Phi_{ij}(t) = 0, \forall t>0$. This gives 
\begin{align}
\frac{\text{d}\Phi_{ij}(t)}{\text{d} t} = \frac{\partial \Phi_{ij}(t)}{\partial p_i} \dot p_i +   \frac{\partial \Phi_{ij}(t)}{\partial p_j} \dot p_j  + \frac{\partial \Phi_{ij}(t)}{\partial t} = 0
\end{align}
Therefore, to enforce the equality constraint, vehicle $j$'s motion should satisfy
\begin{align}
   \frac{\partial \Phi_{ij}(t)}{\partial p_j} \dot p_j  = - \frac{\partial \Phi_{ij}(t)}{\partial p_i} \dot p_i  - \frac{\partial \Phi_{ij}(t)}{\partial t}    
\end{align}

\begin{remark}
If one assumes a time-invariant equality constraint function $\Phi_{ij}$, then $\frac{\partial \Phi_{ij}(t)}{\partial t} = 0$ and the above condition simplifies to 
\begin{align}
    \frac{\partial \Phi_{ij}(t)}{\partial p_j} \dot p_j  = - \frac{\partial \Phi_{ij}(t)}{\partial p_i} \dot p_i
\end{align}
\end{remark}

Now we further consider the inequality constraint $\mathcal{I}_{ij}(p_i, p_j)$ associated with the edge $(i, j)$, while the follower vehicle $j$ is responsible to take care of the inequality constraint $\mathcal{I}_{ij}(p_i, p_j) \leq 0$. Suppose at time $t$ the inequality constraint is active in the sense that $\mathcal{I}_{ij}(p_i, p_j) = 0$. By the viability theory and set-invariance control, vehicle $j$'s motion should satisfy 
\begin{align}
\frac{\text{d}\mathcal{I}_{ij}(t)}{\text{d} t} = \frac{\partial \mathcal{I}_{ij}(t)}{\partial p_i} \dot p_i +   \frac{\partial \mathcal{I}_{ij}(t)}{\partial p_j} \dot p_j  \leq 0
\end{align}
or equivalently 
\begin{align}
 \frac{\partial I_{ij}(t)}{\partial p_j} \dot p_j \leq -   \frac{\partial I_{ij}(t)}{\partial p_i} \dot p_i   
\end{align}

Further note that vehicle $j$'s motion is subject to the kinematics constraint in \eqref{eq:dynamics_constraint_transformation}
 \begin{align} 
 \Omega_{K_j}  \dot p_j = T_{K_j}.
\end{align}

To summarize, the condition for feasible coordination for a leader-follower vehicle team is stated as follows.
\begin{theorem}
The coordination task for a leader-follower vehicle team with both equality and inequality constraints has feasible motions if, for all follower vehicles $i = 1,2, \cdots, n$, the following \textit{mixed} (in)equalities have solutions
\begin{align}
\Omega_{K_j}  \dot p_j &= T_{K_j} \nonumber \\
\frac{\partial \Phi_{ij}(t)}{\partial p_j} \dot p_j  &= - \frac{\partial \Phi_{ij}(t)}{\partial p_i} \dot p_i  - \frac{\partial \Phi_{ij}(t)}{\partial t}  \nonumber \\
\frac{\partial I_{ij}(t)}{\partial p_j} \dot p_j &\leq -   \frac{\partial I_{ij}(t)}{\partial p_i} \dot p_i,\,\,\, \text{if}\, (i, j) \in \chi(P)
\end{align} 
where $\chi(P)$ denotes the set of active constraints among all the edges. 
\end{theorem}

To determine feasibility of the coordination task for the whole team, by following Algorithm \ref{algorithm:feasibility_undirected}, a recursive procedure can be performed to all the vehicles in the leader-follower group, starting from the top leader to the last follower in the underlying tree graph. We note in contrast to the undirected graph case,   such a recursive procedure for the directed tree graph for a leader-follower team enables a \textit{decentralized}  checking of the feasibility condition for each vehicle where the codistribution matrix $\Omega_K, \Omega_E, \Omega_I$ only involves vehicle~$j$ and the associated edge $(i, j)$ in Algorithm~\ref{algorithm:feasibility_undirected}, and the procedure can be terminated within a finite step. 

\section{Case study: coordinating multiple vehicles with distance and heading constraints} \label{sec:case_study}

\subsection{Typical vehicle kinematics}
In this section, we consider several application examples with case studies to illustrate the proposed coordination theory and algorithms. These application examples involve the coordination of homogeneous or heterogeneous vehicles subject to various combinations of constraints. We consider three types of vehicles: a unicycle-type vehicle, a constant-speed vehicle and a car-like vehicle.

The unicycle vehicle is described by \begin{align} \label{eq:example_unicycle}
    \dot x_i &= v_i \,\, \text{cos}(\theta_i), \nonumber \\
\dot y_i &= v_i \,\, \text{sin}(\theta_i),   \\
\dot \theta_i &=   u_i,  \nonumber
\end{align}
where the state variable is $p_i = [x_i, y_i, \theta_i]^{\top} \in \mathbb{R}^2 \times \mathbb{S}^1 \in \mathbb{R}^3$. 
The kinematic constraint for a unicycle-type vehicle can be equivalently stated by the annihilating codistribution
\begin{align}
\Omega_{K_i} = \Delta_i^\perp = \text{span}\{\text{sin}(\theta_i) \text{d}x_i-\text{cos}(\theta_i) \text{d}y_i\}.
\end{align}
Now consider a nonholonomic vehicle with \textit{constant-speed constraints}, which can also be described by \eqref{eq:example_unicycle} but with a \textit{fixed} speed $v_i$.  The only control input is $u_i$ that steers the vehicle's orientations. 
Introducing the two vector fields  
$
f_{i,0} = \left[
v_i\text{cos}(\theta_i),
v_i\text{sin}(\theta_i),
0
\right]^\top,
f_{i,1} = \left[
0,
0,
1
\right]^\top
$,
we can rewrite the constant-speed vehicle model as
\begin{equation} \label{eq:constant_speed_model}
\dot p_i = [\dot x_i, \dot y_i, \dot \theta_i]^{\top} = f_{i,0} + f_{i,1} u_{i}.
\end{equation}
Denote the  two linearly independent covectors  of   the codistribution   as $\omega_{i,1}$ and $\omega_{i,2}$. 
With the dual vector basis $(\text{d}x_i, \text{d}y_i, \text{d}\theta_i)$, one can show an explicit expression of the covectors
$
\omega_{i,1} =  \text{sin}(\theta_i) \text{d}x_i - \text{cos}(\theta_i)\text{d}y_i, \, 
 \omega_{i,2} =  \text{cos}(\theta_i)\text{d}x_i+\text{sin}(\theta_i)\text{d}y_i \nonumber
$ (see \cite{Sun2016feasibility}). 
The affine codistribution is obtained as $\Omega_{K,i} = [
\omega_{i,1}^{\top},
\omega_{i,2}^{\top}
]^{\top}$, and there holds   $\Omega_{K,i}f_{i,1} = 0$ and $\Omega_{K,i}f_{i,0} = T_{K_i}$, where $T_{K_i} = [q_{i,1}, q_{i,2}]^{\top} = [0 , v_i]^{\top}$. 

Further consider a car-like vehicle, whose kinematic equation is described by (see \cite{de1998feedback})
\begin{align} \label{eq:car_model}
\dot x_i &= u_{i,1}   \text{cos}(\theta_i), \nonumber \\
\dot y_i &= u_{i,1}  \text{sin}(\theta_i),   \nonumber \\
\dot \theta_i &= u_{i,1} (1/l_i)  \text{tan} (\phi_i),    \nonumber \\
\dot \phi_i &=  u_{i,2},
\end{align}
with the state variables $p_i = (x_i, y_i, \theta_i, \phi_i) \in \mathbb{R}^2 \times \mathbb{S}^1 \times \mathbb{S}^1$, where $(x_i, y_i)$ are the Cartesian coordinates of the rear wheel, $\theta_i$ is the orientation angle of the vehicle
body with respect to the $x$ axis,  $\phi_i$ is the steering angle, and $l_i$ is the distance between the midpoints of the two wheels. The model \eqref{eq:car_model} describes kinematic motions for a typical rear-wheel-driving car, which is subject to two non-holonomic motion constraints (rolling without slipping sideways for each wheel, respectively). 
In an equivalent compact form, one can write
\begin{align}
    \dot p_i = [\dot x_i, \dot y_i, \dot \theta_i, \dot \phi_i]^{\top} = f_{i,1} u_{i,1} + f_{i,2} u_{i,2},
\end{align}
with $f_{i,1} = [
\text{cos}(\theta_i),
\text{sin}(\theta_i),
(1/l_i)  \text{tan} (\phi_i),
0
]^\top$ and $f_{i,2} = [
0,
0,
0,
1
]^\top$.
 The distribution generated by the two vector fields $f_{i,1}$ and $f_{i,2}$ is described by $\Delta_i = \text{span}\{f_{i,1}, f_{i,2}\}$, which can be equivalently stated by the annihilating co-distribution:
$\Omega_{K_i} = \Delta_i^\perp = \text{span}\{\text{sin}(\theta_i+ \phi_i) \text{d}x_i-\text{cos}(\theta_i+\phi_i) \text{d}y_i - l_i \text{cos}(\phi_i)\text{d}\theta_i, \text{sin}(\theta_i) \text{d}x_i-\text{cos}(\theta_i) \text{d}y_i\}$. 
\begin{figure}[t]
\begin{center}
\vspace{-10pt}
\includegraphics[width=0.5\textwidth]{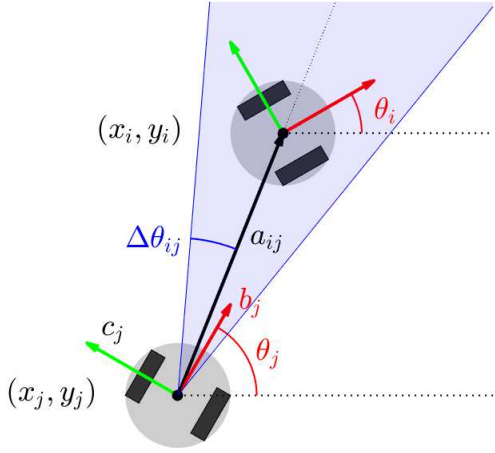}
\vspace{-40pt}
\caption{Illustration of a visibility inequality constraint, $\mathcal{I}^{(4)}_{ij}$, bounding the direction $b_j$ to the blue cone defined by $a_{ij}$ and the angle $\Delta\theta_{ij}$.}
\label{fig:geometry}
\end{center}
\end{figure}

\subsection{Modelling of vehicle coordination constraints} \label{sec:modelling_constraints}
Consider two of the previously defined vehicles in the form~\eqref{eq:dynamics_constraint_transformation} sub-indexed $i$ and $j$ respectively, as illustrated in Figure~\ref{fig:geometry}. A common coordination task may include a simple inter-vehicle distance constraint, with 
\begin{align}\label{eq:distanceconstraint}
    \Phi_{ij}^{(1)}:  \frac{1}{2}(x_i - x_j)^2 +  \frac{1}{2}(y_i - y_j)^2  -  \frac{1}{2}d_{ij}^2 = 0,
\end{align}
for some $d_{ij}>0$, which generates a codistribution matrix
\begin{equation}
\Omega_{E, ij}^{(1)} = [(x_i - x_j)(\text{d}x_i - \text{d}x_j) + (y_i - y_j)(\text{d}y_i - \text{d}y_j)].
\end{equation}
Practical coordination tasks may also include a distance constraint in terms of a two-sided inequality,
\begin{small}
\begin{align}  \label{eq:distance_inequality_constraint}
    \mathcal{I}_{ij}^{(1)}:  \frac{1}{2}(d_{ij}^{-})^2\leq  \frac{1}{2}(x_i - x_j)^2 +  \frac{1}{2}(y_i - y_j)^2 \leq  \frac{1}{2}(d_{ij}^{+})^2,  
\end{align}
\end{small}
with $d_{ij}^{-},d_{ij}^{+}>0$, and codistribution matrix given by $\Omega_{I, ij}^{(1)}=\Omega_{E, ij}^{(1)}$ if the right inequality becomes active, or   $\Omega_{I, ij}^{(1)}= -\Omega_{E, ij}^{(1)}$ if the left inequality becomes active. 

Some tasks may require heading constraints in the form
\begin{align} 
    \Phi_{ij}^{(2)}: \theta_i - \theta_j = \delta_{ij},
\end{align}
for some constant $\delta_{ij}>0$. The corresponding codistribution of this constraint takes the form $\Omega_{E, ij}^{(2)} = [\theta_i \text{d} \theta_i - \theta_j \text{d}\theta_j]$. Similarly to the distance inequality constraints, we define a closely related constraint,
\begin{align}\label{eq:heading_inequality_constraint1}
    \mathcal{I}_{ij}^{(2)}: \delta^-_{ij} \leq  \theta_i - \theta_j \leq \delta_{ij}^+.
\end{align}
The heading inequality constraint in \eqref{eq:heading_inequality_constraint1} generates a codistribution $\Omega_{I, ij}^{(2)}=\Omega_{E, ij}^{(2)}$ if the right inequality in \eqref{eq:heading_inequality_constraint1} becomes active, or $\Omega_{I, ij}^{(2)}= -\Omega_{E, ij}^{(2)}$ if the left inequality in \eqref{eq:heading_inequality_constraint1} becomes active.  When considering tasks of this nature, a more general form of constraint is given by
\begin{align} \label{eq:heading_constraint3}
    \mathcal{I}_{ij}^{(3)}: \Delta\theta_{ij}^{-}\leq\text{arctan}\Big(\frac{y_i-y_j}{x_i - x_j}\Big) - \theta_j \leq \Delta\theta_{ij}^{+},
\end{align}
referred to as a visibility constraint. Such an inequality constraint  has  been used in modelling visibility maintenance control in multi-robotic systems \cite{panagou2014cooperative}. However, the inequality heading constraint in the form of \eqref{eq:heading_constraint3}  suffers by the range of the arctangent function. Consequently, we consider an equivalent inequality constraint, defining the directions $a_{ij} := [x_i - x_j, y_i - y_j]$, $b_{j} := [\cos(\theta_j), \sin(\theta_j)]$, $c_{j} := [-\sin(\theta_j), \cos(\theta_j)]$, and form the equivalent constraint
\begin{align} \label{eq:heading_inequality_constraint3}
    \mathcal{I}_{ij}^{(4)}: \cos(\Delta\theta_{ij})\langle a_{ij}, a_{ij} \rangle^{1/2}\leq  \langle a_{ij},b_{j} \rangle.
\end{align}
By some effort, the associated codistribution can be derived as
\begin{align}\label{eq:headingconst}
\begin{array}{c}
\Omega_{I,ij}^{(4)}\hspace{-1pt}=\hspace{-1pt}\dfrac{\langle a_{ij}, c_j\rangle}{\sqrt{\langle a_{ij}, a_{ij} \rangle}} \Bigg(
\dfrac{1}{\langle a_{ij}, a_{ij} \rangle} \Bigg\langle a_{ij},
\hspace{-3pt}
\Bigg[
\begin{array}{c}
\hspace{-5pt}\text{d}x_i-\text{d}x_j\hspace{-5pt}\\
\hspace{-5pt}\text{d}y_j-\text{d}y_i\hspace{-5pt}
\end{array}
\Bigg]
\hspace{-1pt}
\Bigg\rangle
\hspace{-2pt}+
\hspace{-2pt}\text{d}\theta_j\Bigg).
\end{array}
\end{align}
when the inequality constraint \eqref{eq:heading_inequality_constraint3} becomes active. 
\begin{remark}\label{rem:singA}
It should be noted that the constraint~\eqref{eq:headingconst} may become singular due to the division by $\langle a_{ij}, a_{ij} \rangle$, a corner case to be revisited and addressed in the examples.
\end{remark}

\subsection{Coordinating two unicycle vehicles}
 
In the first example, we consider two unicycle vehicles which are to cooperatively maintain a constant inter-vehicle distance \eqref{eq:distanceconstraint} and a bounded heading displacement or visibility inequality constraint as described above by one of \eqref{eq:heading_inequality_constraint1}, \eqref{eq:heading_constraint3}, or \eqref{eq:heading_inequality_constraint3}. 
Now we construct a joint codistribution matrix from the (non-holonomic) kinematic motion constraints and the distance equality constraint
$$
\Omega =
\left[
\begin{array}{cccccc}
\text{sin}(\theta_1)& - \text{cos}(\theta_1)& 0 &0 &0 &0   \\
0 &0 &0 & \text{sin}(\theta_2) & -\text{cos}(\theta_2)&  0   \\
x_1 - x_2 &y_1 - y_2 & 0     &x_2 - x_1 &y_2 - y_1 &  0
\end{array} \right]
$$
with $T =  [T_K^\top, T_E^\top]^\top = [0, 0, 0]^\top$. Solving the equations $\Omega(\dot P) = T$ yield the solutions $\dot P = \sum_{i=1}^3 w_iK_i$, where
$$
K_1  =
\left[
0,
0,
1,
0,
0,
0   
\right]^\top, \quad K_2  =
\left[
0,
0,
0,
0,
0,
1   
\right]^\top
$$
and
$$
K_3  =
\left[
\begin{array}{c}
\text{cos}(\theta_1)\left(\text{cos}(\theta_2)(x_1 - x_2) + \text{sin}(\theta_2) (y_1 - y_2) \right)   \\
\text{sin}(\theta_1)\left(\text{cos}(\theta_2)(x_1 - x_2) + \text{sin}(\theta_2) (y_1 - y_2) \right)   \\
0   \\
\text{cos}(\theta_2)\left(\text{cos}(\theta_1)(x_1 - x_2) + \text{sin}(\theta_1) (y_1 - y_2) \right)   \\
\text{sin}(\theta_2)\left(\text{cos}(\theta_1)(x_1 - x_2) + \text{sin}(\theta_1) (y_1 - y_2) \right)   \\
0  
\end{array} \right]
$$
It is clear that the virtual controls $w_1$ and $w_2$ generate the angular speeds for each vehicle, respectively, while the term $K_3$ maintains a constant desired distance between them (assuming that initially the distance constraint is met). Furthermore, the solution with $K_1$ and $K_2$ and virtual control inputs $w_1$ and $w_2$ possess the motion freedoms to generate admissible angular input that achieves desired heading re-orientations to satisfy the heading or visibility inequality in the form of  \eqref{eq:heading_inequality_constraint1}-\eqref{eq:heading_inequality_constraint3}. For example, when the heading inequality constraint becomes active in the sense that $\theta_1 - \theta_2 -\delta_{12}^+ = 0$ which renders a codistribution $\Omega_{I, 12}^{(2)}$, any $w_1K_1$ with a negative $w_1$, or any $w_2K_2$ with a positive $w_2$, is a feasible solution guaranteeing $\Omega_{I, 12}^{(2)} \dot P \leq 0$ that generates feasible trajectories for the vehicle group.  The same principle is also applied to other types of heading inequality constraints in the form of \eqref{eq:heading_constraint3} or \eqref{eq:heading_inequality_constraint3}, while feasible motion always exists to ensure the heading or visibility inequality constraint is always satisfied. In summary, we have the following lemma on coordination feasibility and motion generation for two-unicycle vehicle group.

\begin{lemma} \label{lemma:two_vehicle_exm1}
Consider two unicycle-type vehicles, each described by \eqref{eq:example_unicycle},   with a coordination task of maintaining a constant inter-vehicle distance   $d_{12}$ and a bounded heading displacement or visibility inequality constraint. Suppose initially at  time $t = 0$ both constraints are met. By using the above derived control   solutions with the vector functions $K_1, K_2, K_3$:
\begin{itemize}
    \item The distance is preserved by the motion control generated by the derived control with any $w_l$.
    \item If initially the heading/visibility  inequality is satisfied, then a feasible control always exists (with the possible choice of $w_l$) that preserves both distance equality and heading/visibility inequality constraints. 
\end{itemize}
\end{lemma}

\subsection{Coordinating a unicycle and a constant-speed vehicle} \label{sec:constspeedunicycle}
Now we consider a coordination task that involves a constant-speed vehicle and a general unicycle vehicle, aiming to maintain inter-vehicle distance equality and heading inequality constraints for a coordination task. 
The co-distribution matrix from the kinematic equations and equality constraint is constructed by 
$$
\Omega =
\left[
\begin{array}{cccccc}
\text{sin}(\theta_1)& - \text{cos}(\theta_1)& 0 &0 &0 &0   \\
 \text{cos}(\theta_1)&  \text{sin}(\theta_1) &0 &0 &0 &0 \\
0 &0 &0 & \text{sin}(\theta_2) & -\text{cos}(\theta_2)&  0    \\
x_1 - x_2 &y_1 - y_2 & 0     &x_2 - x_1 &y_2 - y_1 &  0
\end{array} \right]
$$
with $T =  [T_K^\top, T_E^\top]^\top = [0, v_1,0,0,0]^\top$. The algebraic equation $\Omega(\dot P) = T$ is solved by,
$$
    \bar K = 
    \left[
\begin{array}{c}
v_1 \text{cos}(\theta_1)   \\
v_1 \text{sin}(\theta_1)   \\
0   \\
\frac{\text{cos}(\theta_2)\left(v_1 \text{cos}(\theta_1)(x_1 - x_2) + v_1 \text{sin}(\theta_1)(y_1 - y_2)\right)}{\text{cos}(\theta_2)(x_1 - x_2) + \text{sin}(\theta_2)(y_1 - y_2)}    \\
\frac{\text{sin}(\theta_2)\left(v_1 \text{cos}(\theta_1)(x_1 - x_2) + v_1 \text{sin}(\theta_1)(y_1 - y_2)\right)}{\text{cos}(\theta_2)(x_1 - x_2) + \text{sin}(\theta_2)(y_1 - y_2)}    \\
0   
\end{array} \right]
$$  and $
K_1  =
\left[
0,
0,
1,
0,
0,
0   
\right]^\top,
K_2  =
\left[
0,
0,
0,
0,
0,
1    \right]^\top
$, which enables an abstraction of the coordination system $\dot P = \bar K + \sum_{l=1}^2 w_l K_l$, and an analysis analogous to Lemma \ref{lemma:two_vehicle_exm1}.
\begin{lemma}
Consider a unicycle-type vehicle and a constant-speed vehicle in a coordination group to maintain inter-vehicle distance equality and heading inequality or visibility constraints described in Section~\ref{sec:modelling_constraints}. By using the above-derived control solutions: 
\begin{itemize}
    \item The distance is preserved with the derived  control vector fields for any $w_1$ and $w_2$. 
    \item If initially the heading or visibility inequality is satisfied, then a feasible motion always exists  with possible $w_1$ and $w_2$ that preserves both distance equality and heading/visibility inequality constraints. 
\end{itemize}
\end{lemma}

\begin{remark}\label{rem:singB}
Note that there always exists a direction
$\langle a_{12}, b_2 \rangle = \text{cos}(\theta_2)(x_1 - x_2) + \text{sin}(\theta_2)(y_1 - y_2) = 0$ which makes the solution singular. In practice, this caveat can be solved by imposing additional constraints on $\langle a_{12}, b_2 \rangle$.
\end{remark}

\subsection{Coordinating a unicycle and a car-like vehicle}
Consider a two-vehicle group, one described by the uni-cycle equation and the other by a car-like dynamics. The two vehicles assume a task to cooperatively maintain a constant distance $d_{12}$ and a heading or visibility inequality constraint.

The joint codistribution matrix from both kinematic constraint and distance equality constraint can be obtained as (using the dual space bases $[\text{d}x_1,\text{d}x_2, \cdots, \text{d}\phi_2,\text{d}\theta_2]$): $\Omega = [\text{sin}(\theta_1) \text{d}x_1-\text{cos}(\theta_1) \text{d}y_1, \text{sin}(\theta_2+ \phi_2) \text{d}x_2-\text{cos}(\theta_2+\phi_2) \text{d}y_2 - l_2 \text{cos}(\phi_2)\text{d}\theta_2, \text{sin}(\theta_2) \text{d}x_2-\text{cos}(\theta_2) \text{d}y_2, (x_1 - x_2)(\text{d}x_1 - \text{d}x_2) + (y_1 - y_2)(\text{d}y_1 - \text{d}y_2)]$.

The solution to the algebraic equation $\Omega(\dot P) = T =  0$ is obtained as
$
    \dot P = \sum_{l=1}^3 w_l K_l
$ with 
$
K_1  =
\left[
0,
0,
1,
0,
0,
0,
0   
  \right]^\top,
K_2  =
\left[
0,
0,
0,
0,
0,
0,
1   
  \right]^\top  
  $, and $$
K_3  =
\left[
\begin{array}{c}
\text{cos}(\theta_1)\left(\text{cos}(\theta_2)(x_1 - x_2) + \text{sin}(\theta_2) (y_1 - y_2) \right)   \\
\text{sin}(\theta_1)\left(\text{cos}(\theta_2)(x_1 - x_2) + \text{sin}(\theta_2) (y_1 - y_2) \right)   \\
0   \\
\text{cos}(\theta_2)\left(\text{cos}(\theta_1)(x_1 - x_2) + \text{sin}(\theta_1) (y_1 - y_2) \right)   \\
\text{sin}(\theta_2)\left(\text{cos}(\theta_1)(x_1 - x_2) + \text{sin}(\theta_1) (y_1 - y_2) \right)   \\
\frac{1}{l_2} \text{tan}\phi_2 \left(\text{cos}(\theta_1)(x_1 - x_2) + \text{sin}(\theta_1) (y_1 - y_2) \right)\\
0  
\end{array} \right]
$$
The coordination feasibility and motion generation result is summarized in the following lemma.
\begin{lemma}
Consider a two-vehicle group consisting of a  unicycle-type vehicle and a car-like vehicle, with a coordination task of maintaining a constant inter-vehicle distance   $d_{12}$ and a   heading/visibility constraint described in Section~\ref{sec:modelling_constraints}. Suppose initially at  time $t = 0$ both constraints are met.  By using the above derived control solutions: 
\begin{itemize}
    \item The inter-vehicle distance is preserved with the above-derived control for any $w_l$. 
    \item If initially the heading/visibility inequality is satisfied, then a feasible control always exists (with the possible choice of $w_l$) that preserves both distance equality and heading/visibility inequality constraints. 
\end{itemize}
\end{lemma}

\subsection{Multiple homogeneous vehicles with mixed constraints}\label{sec:multihomogenous}
Now we consider a leader-follower vehicle group with mixed constraints. Consider multiple unicycle models described by~\eqref{eq:example_unicycle}, with one leader vehicle $p_1(t)\in \mathbb{R}^3$ and two followers $p_2(t),p_3(t)\in \mathbb{R}^3$. The kinematics yield an annihilating co-distribution $\sin(\theta_i)\text{d}x_i - \cos(\theta_i)\text{d}y_i = 0$, resulting in an $\Omega_K\in \mathbb{R}^{3\times 9}$ with $T_K = [0,0,0]^{\top}$. The leader is constrained to follow an arbitrary reference trajectory in terms of two continuous control inputs $v_{1,r}(t), u_{1,r}(t)\in C^0$. Much like the example in Section~\ref{sec:constspeedunicycle}, these time-varying speeds are  incorporated as two equality constraints, with
$$
\cos(\theta_1)\text{d}x_1 + \sin(\theta_1)\text{d}y_1 = v_{1,r}(t), \quad \text{d}\theta_1 = u_{1,r}(t),
$$
represented in the standard compact matrix form with $\Omega_E\in \mathbb{R}^{2\times 9}$ with $T_E = [v_{1,r}(t), u_{1,r}(t)]^{\top}\in \mathbb{R}^2$.

\begin{figure}[t]
\begin{center}
\includegraphics[width=0.5\textwidth]{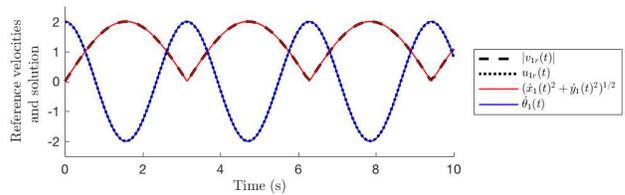}
\caption{Reference trajectory for the leader vehicle $[v_{1r}(t),u_{1r}(t)]^{\top}$ (black), with the speeds $(\dot{x}_1(t)^2+\dot{y}_1(t)^2)^{1/2}$ (red) and $\dot{\theta}_1(t)$ (blue) found in the computed solution.}
\label{fig:simulation_example1}
\end{center}
\end{figure}

In order for the followers to maintain visibility of the leader, we pose two inequality constraints in the form~\eqref{eq:heading_inequality_constraint3}, enforcing $\mathcal{I}_{12}^{(4)}$ and $\mathcal{I}_{13}^{(4)}$ with a maximum heading angle of $\Delta \theta_{12} = \Delta \theta_{13} = 0.4$ (rad). The annihilating co-distributions $\Omega_{I,12}^{(4)}$ and  $\Omega_{I,13}^{(4)}$ are given in~\eqref{eq:headingconst}, which are omitted here for brevity. As was noted in Remark~\ref{rem:singA}, these distributions exist when the distance between the vehicles is non-zero. To eliminate the possibility of singular solutions, a distance inequality constraint is posed in the form~\eqref{eq:distance_inequality_constraint} as $\mathcal{I}_{12}^{(1)}$ and $\mathcal{I}_{13}^{(1)}$, with $d_{12}^-=d_{13}^-=1$ and $d_{12}^+=d_{13}^+=2$. In Remark~\ref{rem:singB}, we noted that there exists a direction $\langle a_{1j}, b_j \rangle=0$, at which the motion solution becomes singular when activating any distance constraint $\mathcal{I}^{(1)}_{1j}$. This caveat is conveniently avoided by the posed heading inequality constraint, effectively enforcing bounding $\langle a_{1j}, b_j \rangle \geq d_{1j}^-\cos(\Delta\theta_{1j})=0.92$. Consequently, any feasible motion found by Algorithm~\ref{algorithm:feasibility_undirected} satisfying the posed inequality constraints gives rise to non-singular, well-defined solution control flows. Combining the constraints yields
$$
\Omega_I = [
(\Omega_{I,12}^{(1)})^{\top}
(\Omega_{I,13}^{(1)})^{\top}
(\Omega_{I,12}^{(4)})^{\top}
(\Omega_{I,13}^{(4)})^{\top}
]\in \mathbb{R}^{6\times 9},
$$
of which at most four constraints may be active at any point in time (the distance upper and lower bound cannot be met simultaneously). This complex system with one leader vehicle (with predefined constrained speeds) and two following unicycles always has feasible coordination motions in all possible combinations of these constraints when checked with Algorithm~\ref{algorithm:feasibility_undirected}. To show the found solutions in practice, a simulation was run with the three vehicles, recomputing the virtual inputs $w_i \in \mathbb{R}$ each time an inequality  constraint was activated. We consider a leader vehicle reference trajectory
$$
v_{1r}(t) = 2\sin(t), \quad u_{1r}(t) = 2\cos(2t),
$$
which is followed perfectly when incorporated through time-varying equality constraints, as demonstrated in   Figure~\ref{fig:simulation_example1}.

\begin{figure}[t]
\begin{center}
\includegraphics[width=0.5\textwidth]{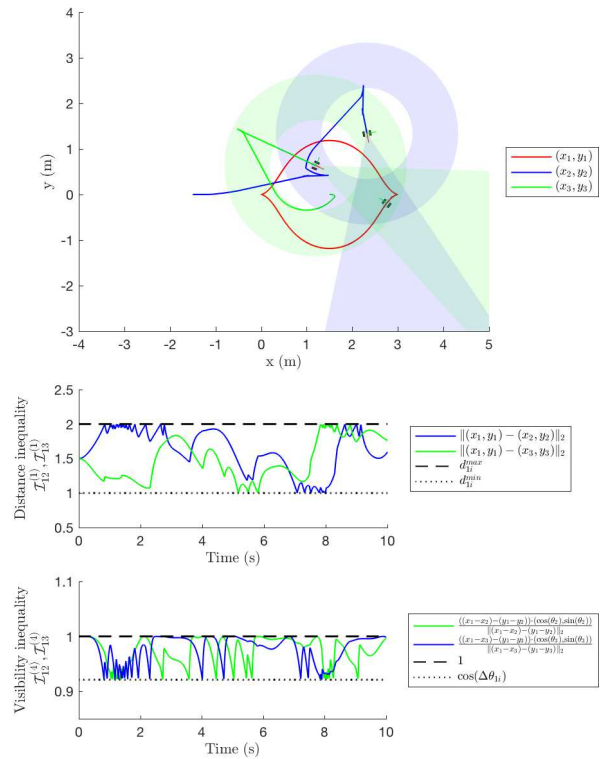}
\vspace{-40pt}
\caption{\textit{Top}: Two-dimensional plot of the three unicycle vehicles, with the leader (red) and the two followers (blue and green). \textit{Center}: Inequality-constrained distance between the leader and the followers. \textit{Bottom}: Inequality-constrained cosine angle $\langle c_{1j}, b_j \rangle$ for visibility maintenance between the leader and the followers.}
\label{fig:simulation_example2}
\end{center}
\end{figure}

Furthermore, the two-dimensional trajectories of the leader (red) and followers (blue, green) are depicted in Figure~\ref{fig:simulation_example2}, along with the distance inequality constraints $\{\mathcal{I}_{12}^{(1)},\mathcal{I}_{13}^{(1)}\}$, and the cosine angle inequality constraints $\{\mathcal{I}_{12}^{(4)},\mathcal{I}_{13}^{(4)}\}$ for maintaining visibility. All inequality constraints are met at all times. 


\subsection{Two heterogeneous vehicles with mixed constraints}
\begin{figure}[t]
\begin{center}
\includegraphics[width=0.5\textwidth]{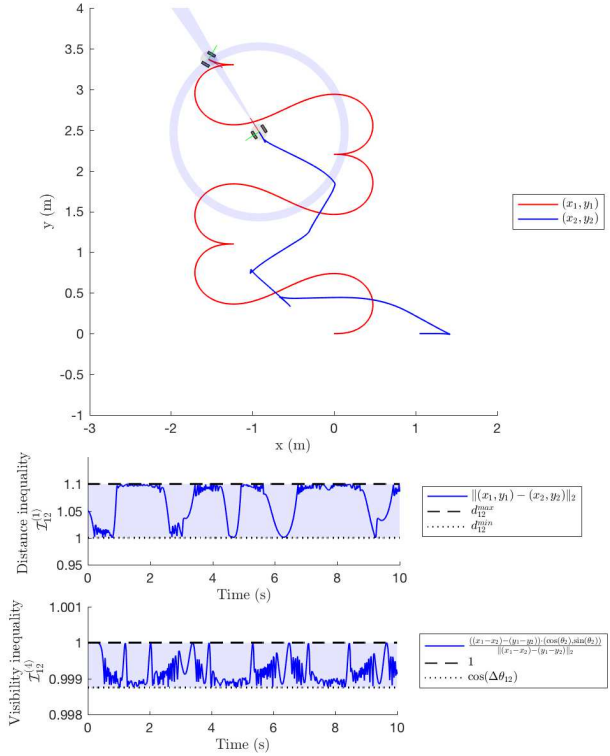}
\caption{Solution trajectories in the positional domain (top), with the distance inequality constraint (center) and the visibility inequality constraint (bottom).}
\label{fig:simulation_example3}
\end{center}
\end{figure}

To show the versatility of the theory, we give a final example with a car-like model with states $p_1 = [x_1,y_1,\theta_1,\phi_1]^{\top}$ with the kinematics in~\eqref{eq:car_model} and a unicycle vehicle defined by the states $p_2 = [x_2,y_2,\theta_2]^{\top}$ in~\eqref{eq:example_unicycle}. Similar to the previous example in Section~\ref{sec:multihomogenous}, we form the matrix $\Omega_K\in \mathbb{R}^{3\times 7}$ with $T_K = [0, 0, 0]^{\top}$ and constrain the car-like vehicle speeds with additional equality constraints in order for it to follow a predefined trajectory. We consider this trajectory in terms of the controls  $u_{11,r}(t), u_{12,r}(t)\in C^0$, and enforce it through the equality constraints given by the affine codistributions,
$$
\cos(\theta_1)\text{d}x_1 + \sin(\theta_1)\text{d}y_1 = u_{11,r}(t), \quad \text{d}\phi_1 = u_{12,r}(t),
$$
which may be represented in a compact matrix form with $\Omega_E\in \mathbb{R}^{2\times 7}$ with $T_E = [u_{11,r}(t), u_{12,r}(t)]^{\top}\in \mathbb{R}^2$. In addition to the equality constraints, we pose a distance inequality constraint $\mathcal{I}_{12}^{(1)}$ with very narrow bounds, $d_{12}^-=1$ and $d_{12}^+=1.1$ (m), and a visibility constraint with a very tight angle bound $\Delta\theta_{12}=0.05$ (rad). Combined, the constraints defined an extremely narrow feasible region with,
$$
\cos(\Delta\theta_{12})=0.998 \leq \frac{\langle a_{12}, b_{2}\rangle}{\langle a_{12}, a_{12}\rangle} \leq 1.
$$
When implementing the reference trajectory of
$$
u_{11,r}(t) = 2\sin(t), \quad u_{12,r}(t) = 2\cos(2t)
$$
and parameterizing the car model with $l_1=0.5$, the found solution and vehicles' trajectories are depicted in Figure~\ref{fig:simulation_example3}. The positional trajectory of the reference vehicle differs greatly from the previous example, due to the implementation of the car-like vehicle kinematics instead of the unicycle kinematics for the leading vehicle (red). We note that the vector fields switch frequently, as the inequality constraints activate often requiring new values of $w_i$ to be computed by the Algorithm~\ref{algorithm:feasibility_undirected}. Nonetheless, the found solution satisfies the kinematic constraints, the equality constraints for the reference trajectory following and the posed inequality constraints in terms of the distance, $\mathcal{I}_{12}^{(1)}$, and visibility,  $\mathcal{I}_{12}^{(4)}$, clearly visible in the lower two plots of Figure~\ref{fig:simulation_example3}.

\section{Conclusions} \label{sec:conclusions}
In this paper, we discuss the coordination control problem for multiple mobile vehicles subject to various constraints (nonholonomic motion constraints, holonomic formation constraints, equality or inequality constraints, among others). Using tools from differential geometry, distribution/codistributions for control-affine systems and viability theory, we have developed a general framework to determine whether feasible motions exist for a multi-vehicle group that meet both kinematic constraints and coordination constraints with a mix of inequality and equality functions for describing a coordination task. A heuristic algorithm is proposed to find feasible motions and trajectories for a group of homogeneous or heterogeneous vehicles to achieve a coordination task. We also provide several case study examples and simulation experiments to illustrate the proposed coordination control schemes.

\bibliography{Feasibility}
\bibliographystyle{ieeetr}

\end{document}